\documentclass[fleqn,twoside]{article}
\usepackage{espcrc2}

\usepackage{graphicx}
\newcommand{\hmu}{\hat{\mu}}
\newcommand{\hrho}{\hat{\rho}}
\def\tr{{\rm ~tr}\,}

\title{Fermionic Symmetry in Ichimatsu-Decomposed Lattice Models}

\author{K. Itoh\address{Faculty of Education, 
        Niigata University, Niigata 950-2181, Japan},
        M. Kato\address{Institute of Physics, University of Tokyo,
                        Komaba, Meguroku, Tokyo 153-8902, Japan},
        H. Sawanaka\address[DPNU]{Department of Physics, Niigata University, 
        Niigata 950-2181, Japan}\thanks{Talk presented by H. Sawanaka.},
        H. So\addressmark[DPNU]
        and
        N. Ukita\addressmark[DPNU]}

\begin{document}

\begin{abstract}
We present the lattice models with exact fermionic symmetries relating
fermions and link variables. 
The plaquettes are distributed in an Ichimatsu pattern (chequered).
% (chessboard) pattern. 
We explain this peculiar structure allows us to have a translation 
in the algebra of the fermionic symmetries.
\vspace{1pc}
\end{abstract}

\maketitle

%%%%%%%%%%%%%%%%%%%%%%%%%%%%%%%%%%%%%%%%%%%%%%%%%%%%%%%%%%%%%%%%%%%%%%
%%%%%%%%%%%%%%%%%%%%%%%%%%%%%%%%%%%%%%%%%%%%%%%%%%%%%%%%%%%%%%%%%%%%%%
\section{Exact fermionic symmetry on the lattice}

To study dynamics of super Yang-Mills theory,
a lattice formulation is an attractive tool.
There are some approaches,
however limited, not exact, realization has done \cite{c-v,k-s}.

Why do we stress the exact symmetry?
It is because the point of our approach,
the scenario is that the exact fermionic symmetry on lattice with
keeping $O(1)$ symmetry in the continuum limit,
then it's just SUSY if it is not BRS symmetry.
So, the first step of our trial is to construct exact fermionic,
not BRS-like symmetry on lattice.

We present new lattice models with an exact fermionic symmetry.
We start with considering in the system with minimal degrees of freedom,
realized as one-cell model,
derive the fermionic transformation, which is expected to
contain the continuum supersymmetry.
Then it is extended to the entire lattice in a non-trivial way,
it is multi-cell model.
We perform it by introducing the particular lattice structure
with a special pattern called `Ichimatsu' pattern in Japanese.
Also using that pattern, we give another model, which we call pipe model.

Finally, we reach cell and pipe mixed model.
%%%%%%%%%%%%%%%%%%%%%%%%%%%%%%%%%%%%%%%%%%%%%%%%%%%%%%%%%%%%%%%%%%%%%%

\section{Our models}

First, we consider a fundamental lattice.
Let us refer it as a cell.
The action consists of gauge and fermion part :
\begin{equation}
 S_g = -\beta \sum_{n,\mu\nu} \tr 
       \left( U_{n,\mu\nu} + U_{n,\nu\mu} \right) \ \ ,
\end{equation}
\begin{equation}
 S_f = \sum_{n,\rho} b_{\rho}(n) \tr \left(
       \xi_{n} U_{n,\rho} \xi_{n+\hrho} U_{n,\rho}^{\dagger} \right) \ \ .
\end{equation}
The gauge action $S_g$ is the plaquette action,
here $\beta$ is the gauge coupling constant.
For the fermion part, we put real staggered fermion on each site.
The fermion action $S_f$ consists of terms which corresponds to
fermions located at neighbouring sites connected by link variables.
The coefficient $b_{\rho}(n)$ is a sign factor for the staggered fermion.

We introduce a fermionic transformation for
which mixes the fermion and the link variables.
We assume the form
to realize the ordinary SUSY transformation in a continuum limit.
Let us call our transformation as pre-SUSY transformation.
The SUSY transformation of a fermion is to give the field strength,
$\delta\psi \sim F_{\mu\nu}\gamma_{\mu\nu}\epsilon$,
hence, 
our ansatz of the fermion transformation is given as
\begin{equation}%\label{equ:dxi}
 \delta \xi_{n} = \sum_{0<\mu<\nu} C_{n,\mu\nu}
                  \left( U_{n,\mu\nu} - U_{n,\nu\mu}\right)
\end{equation}
where $C_{n,\mu\nu}$ is a Grassmann-odd parameter.
We chose the transformation of the link variables in order
the operator from fermion action under the fermion transformation,
to balance the one which produced from the change of the gauge action.
\begin{eqnarray*}
 \delta U_{n,\mu}=& \sum_{\rho}\left\{\left(
 \alpha_{n,\mu}^{\rho} U_{n,\rho}\xi_{n+\hrho}U_{n+\hrho,\rho}^{\dagger}
 \right) U_{n,\mu} \right. \\
 & \!\!\!\!\!\!\!\!\! \left. +~U_{n,\mu}\left(\alpha_{n+\hmu,\mu}^{\rho}
 U_{n+\hmu,\rho}\xi_{n+\hmu+\hrho}U_{n+\hmu,\rho}^{\dagger}\right)
 \right\}
\end{eqnarray*}
where $\alpha$ is a Grassmann-odd parameter.

Now we are going to check
the action invariance under pre-SUSY transformation.
The variation of the action under the pre-SUSY
contains terms cubic and linear in the fermion variables.
For the action to be invariant,
these two sets of terms should vanish separately:
\begin{equation}
 \delta S =0 \Rightarrow
 \left\{\begin{array}{l}
 \delta_{U} S_f = 0 \\ \delta_{U} S_g + \delta_{\xi} S_f = 0
 \end{array}\right. \ \ ,
\end{equation}
corresponding to these conditions,
we obtain two types of relations between introduced parameter:

\begin{equation}
 b_{\mu}(n)\alpha_{n,\mu}^{\rho} + b_{\rho}(n)\alpha_{n,\rho}^{\mu}=0 \ \ ,
\end{equation}

\begin{equation}
 b_{\rho}(n) C_{n,\mu\nu} \sim \beta 
 \left(\alpha_{n,\mu}^{\rho} - \alpha_{n,\nu}^{\rho}\right) \ \ .
\end{equation}

We can check out exact symmetry in this one-cell model.
Solving these relations,
we find there are $D-1$ independent parameters for each site.

%%%%%%%%%%%%%%%%%%%%%%%%%%%%%%%%%%%
Then, we consider our model with pre-SUSY in an entire lattice space.
Since naive extension only produces $O(a)$ symmetry,
to be discussed later,
here we take a sophisticated way.
We put a restriction on the plaquette variables.
In the simplest 2-dimensional space,
the allowed plaquette variables form `Ichimatsu' pattern as
%%%%%%%%%%%%%%%%%%%%%%%%%%%%%%%%%%%%%%%%
%%%%%%%%%%%%%%%%%%%%%%%%%%%%%%%%%%%%%%%%
\begin{figure}[h]
 \includegraphics[scale=0.7]{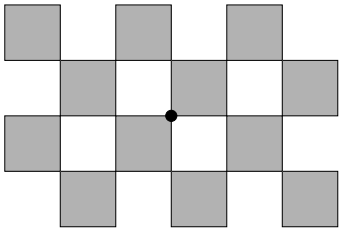}
 \caption{2-dimensional Ichimatsu pattern.}
 \label{2d-ichimatsu}
\end{figure}
%%%%%%%%%%%%%%%%%%%%%%%%%%%%%%%%%%%%%%%%
%%%%%%%%%%%%%%%%%%%%%%%%%%%%%%%%%%%%%%%%
Fig. \ref{2d-ichimatsu}.
Contrary to the ordinary lattice theory,
not all the possible plaquette variables are included in this extension.
All the fermion and link variables are included,
while only the plaquette variables are restricted.
Even in arbitrary space dimension,
we can do this multi-cell extension in the way that
there presents the same pattern on any two-dimensional surfaces.

By taking same steps in the one-cell case,
one can find that the action invariance under pre-SUSY transformation
in multi-cell included model with the Ichimatsu pattern.
However, it is easy to suppose that
in the cell model, the perturbative picture is not present,
since the interaction between cells is introduced through the fermion
variables, which belong to two neighbouring cells.
(Details are reported in our previous paper \cite{ikssu1}.)

%%%%%%%%%%%%%%%%%%%%%%%%%%%%%%%%%%%
By the way, here we construct another model,
by exchanging the allowed and discarded plaquette variables,
since there are two ways to put the Ichimatsu pattern
on a two-dimensional surface.
Now let us call this alternative model, pipe model,
from its structure in 3-dim. case.

The pipe model is present for any dimension,
(anyhow this distinction is meaningless in 2-dim.)
its plaquette variables are arranged in the complimentary to the cell
model.
It may say that we decomposed ordinary lattice space
using the Ichimatsu pattern.
%%%%%%%%%%%%%%%%%%%%%%%%%%%%%%%%%%%%%%%%
%%%%%%%%%%%%%%%%%%%%%%%%%%%%%%%%%%%%%%%%
\begin{figure}[h]
 \includegraphics[scale=0.9]{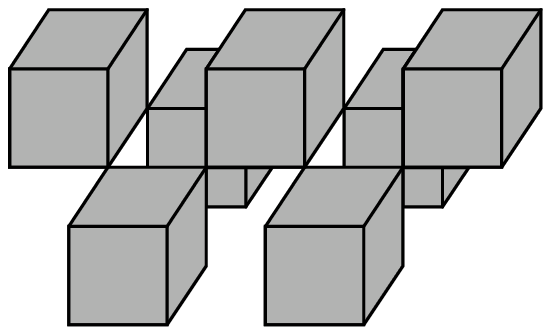}
\vspace{5pt} \\ 
 \includegraphics[scale=0.8]{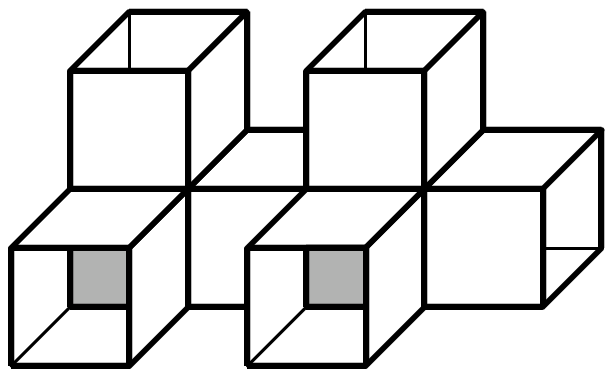} 
 \caption{
3-dimensional lattice structure of multi-cell and pipe models.
These two models are complementary to each other.}
\end{figure}
%%%%%%%%%%%%%%%%%%%%%%%%%%%%%%%%%%%%%%%%
%%%%%%%%%%%%%%%%%%%%%%%%%%%%%%%%%%%%%%%%

%%%%%%%%%%%%%%%%%%%%%%%%%%%%%%%%%%%%%%%%%%%%%%%%%%%%%%%%%%%%%%%%%%%%%%
This model differs from the cell model only in the gauge sector.
However, repeating the same procedure as the cell model,
we find the presence of the pre-SUSY in the pipe model.

We do not encounter the difficulty in the cell model
even when we remove the fermion variables.
However, we may show that the pipe model without fermion variables
do not have an appropriate continuum limit
(can be shown equivalent to trivial model statistical mechanically).

%%%%%%%%%%%%%%%%%%%%%%%%%%%%%%%%%%%
So far, we consider these two models separately.
Now we are going to consider its mixed model to overcome the difficulties.
In defining the mixed model,
we assign different coupling constants to two sets of plaquette
variables in the gauge action,
$\beta_c$ is for those of the cell model and
$\beta_p$ is for the pipe model.

\begin{equation}
 S = S_f + S_{g}^{cell}[\beta_c] + S_{g}^{pipe}[\beta_p]
\end{equation}

While in the pre-SUSY transformation,
the fermion transformation that produces the plaquette variables
is influenced to be producing two sets of plaquette variables:
\begin{eqnarray}\label{equ:dxi}
 \delta \xi_{n}=& 
 \sum_{P_{cell}} C^{cell}_{n,\mu\nu}(U_{n,\mu\nu}-U_{n,\nu\mu})
 \\ & \! +
 \sum_{P_{pipe}} C^{pipe}_{n,\mu\nu}(U_{n,\mu\nu}-U_{n,\nu\mu})
 \ \ , \nonumber 
\end{eqnarray}
where $P_{cell}$ ($P_{pipe}$) means plaquette variables of
the cell (pipe) model.

Although it is not so obvious
whether we may keep the fermionic symmetry or not,
we can find exact fermionic symmetry even in the mixed model.

We note two important features here. 
First, the mixed model may be treated properly even in a perturbative way,
since there included all the plaquette variable
same as in the ordinary lattice.

Second, when we introduced our pre-SUSY transformation,
we assumed the form of the fermion transformation
to be related to the continuum SUSY.
Considering the naive continuum limit of Eq. (\ref{equ:dxi}),
we find it becomes
$$
 \delta \xi_n \! \sim \! (1-\frac{\beta_p}{\beta_c}) \sum_{0<\mu<\nu} \!
 (-)^{n_{\mu}+n_{\nu}}(C^{(+)}_{n,\mu\nu} + C^{(-)}_{n,\mu\nu})F_{n,\mu\nu}
$$
where $C^{(\pm)}_{n,\mu\nu}$ are Grassmann-odd parameters
related to the parameter $C$.
It is proportional to the field strength,
but is also proportional to the difference of the two coupling constants
$\beta_c$ and $\beta_p$.
This shows two couplings, $\beta_c$ and $\beta_p$ must differ
to have proper continuum limit,
is the reason for the transformation will be higher order and vanish in
the limit in the ordinary lattice model.

If the fermionic symmetry is really related to the expected SUSY,
the continuum limit is to be studied with the cell and pipe mixed model.
First result of pure gauge system with the Ichimatsu structure is reported
in \cite{ikmss}.

%%%%%%%%%%%%%%%%%%%%%%%%%%%%%%%%%%%%%%%%%%%%%%%%%%%%%%%%%%%%%%%%%%%%%%
\section{Summary}

We presented new lattice models with an exact fermionic symmetry,
as a step towards the super Yang-Mills theory on lattice.

In the last cell and pipe mixed model,
we find the crucial condition towards the continuum SUSY
and
the Ichimatsu pattern plays an important role in our models.
We may confirm that our models satisfy reasonable requirements
for a proper lattice theory,
such as translational and rotational invariance 
and the reflection positivity,
also by the property of the Ichimatsu pattern.

We also work on the realization of the staggered Majorana fermion,
however the complete realization of spinor structure still remains
as an unsolved question.
Now, the recovery of the spinor structure
and the doubling problem are crucially important and most difficult.
Some discussions will be given in our forthcoming paper
\cite{ikssu2}.
%%%%%%%%%%%%%%%%%%%%%%%%%%%%%%%%%%%%%%%%%%%%%%%%%%%%%%%%%%%%%%%%%%%%%%

%%%%%%%%%%%%%%%%%%%%%%%%%%%%%%%%%%%%%%%%%%%%%%%%%%%%%%%%%%%%%%%%%%%%%%
%%%%%%%%%%%%%%%%%%%%%%%%%%%%%%%%%%%%%%%%%%%%%%%%%%%%%%%%%%%%%%%%%%%%%%
\end{document}